\documentclass{pramana}

\usepackage{graphicx,amsmath,bm}

\begin{document}

\title{Chaos suppression in fractional order systems using state-dependent noise}


\author{A. O. Adelakun\textsuperscript{1}, S. T. Ogunjo\textsuperscript{1,*}\and I. A. Fuwape\textsuperscript{1,2}}
\affilOne{\textsuperscript{1} Department of Physics, Federal University of Technology Akure, Ondo State, Nigeria\\}
\affilTwo{\textsuperscript{2} Michael and Cecilia Ibru University, Ughelli North, Delta State, Nigeria.}


\twocolumn[{

\maketitle

\corres{stogunjo@futa.edu.ng}

\msinfo{10 February 2019}{1 April 2019}{10 May 2019}

\begin{abstract}
Noise play a creative role in the evolution of periodic and complex systems which are essential for continuous performance of the system.  The interaction of noise generated within one component of a chaotic system with other component in a linear or nonlinear interaction is crucial for system performance and stability.  These types of noise are inherent, natural and insidious.  This study investigates the effect of state-dependent noise on the bifurcation of two chaotic systems.  Circuit realization of the systems were implemented.  Numerical simulations were carried out to investigate the influence of state dependent noise on the bifurcation structure of the Chen and Arneodo-Coullet fractional order chaotic systems.  Results obtained showed that state dependent noise inhibit the period doubling cascade bifurcation structure of the two systems.  These results poses serious challenges to system reliability of chaotic systems in control design, secure communication and power systems.
\end{abstract}



}]


\doinum{12.3456/s78910-011-012-3}
\artcitid{\#\#\#\#}
\volnum{123}
\year{2016}
\pgrange{1--6}
\setcounter{page}{1}
\lp{6}

\section{Introduction}
Fractional order models are best suited to capture the memory effects in mathematical models \cite{OGUNJO2018451}.  The lack of appropriate methods of solution has made fractional order systems unattractive to scientists.  However, in recent times, adequate mathematical representation and solutions has created a resurgence and renewed interest in the subject of fractional order. The order of fractional derivative is an index of memory \cite{du2013measuring} and as stable as their integer order counterpart \cite{ahmed2007equilibrium}.  Due to their relevance, fractional order models have been considered in finance \cite{chen2008nonlinear}, psychology \cite{ahmad2007fractional} and electronic systems \cite{petras2010fractional}.

Noise is ubiquitous and inherent in nature.  The role of noise in a system depends on how it was introduced and intended function of the system.  Noise is desirable in some systems while it is unwanted in others.  The threshold of chaos can be increased or reduced through noise perturbations \cite{fronzoni1998controlling} and synchronization of chaotic oscillators with noise \cite{astakhov1997synchronization}.  Chaos suppression due to noise has been reported in the Duffing oscillator \cite{chacon1995suppression} , parametrically driven Lorenz system \cite{choe2005chaos}, Belousov-Zhabotinsky reaction \cite{matsumoto1983noise} and logistic map \cite{rajasekar1995controlling}.  Dennis et al \cite{dennis} suggests that noise cannot induce chaos.  However, the introduction of noise has also been found to enhance synchronization of chaotic systems \cite{lin2006using,sanchez} and induce chaos in systems \cite{wei2009noise,zhou2003noise,qin2014random}.  Gao \cite{gao} showed that noise can induced chaos when the noise level is within a narrow range.  It has been posited that real life neurons are robust to noise \cite{carrol}.

The role of different types of noise has been investigated on integer and discrete chaotic systems.  There is the need to understand the dynamic influence of noise on the memory effect of fractional order chaotic systems.  Since memory is an intrinsic property of a system, it is intuitive to understudy the interaction of noise generated within the system states and the system itself. The aim of this study is to investigate the impact of state-dependent noise in a class of fractional order chaotic systems.

\section{Methodology}
\subsection{Fractional order derivatives}
Fractional derivatives has been defined in different ways \cite{podlubny2002analogue}. One of the most common is Riemann-Liouville, which can be denoted as:
\begin{equation}\label{eqn2}
\frac{d^\alpha f(t)}{dt^\alpha}=\frac{1}{\Gamma(n-\alpha)}\frac{d^n}{dt^n}\int^{t}_{0}\frac{f(t)}{(t-\tau)^{\alpha-n+1}}d\tau
\end{equation}
In equation \ref{eqn2}, $\Gamma(.)$ is the gamma function, where $n-1<\alpha<n$ and $n$ is an integer. The laplace transform of equation \ref{eqn2} can be written as
\begin{equation}\label{eqn3}
L\lbrace\frac{d^\alpha f(t)}{dt^\alpha}\rbrace=s^q L\lbrace f(t)\rbrace -\sum^{n-1}_{k=0}s^k\lbrace\frac {d^{\alpha-1-k}f(t)}{dt^{\alpha-1-k}}\rbrace_{t=0}
\end{equation}
If the initial conditions are zero for equation \ref{eqn3}, it implies the Riemann-Liouville fractional derivative is
\begin{equation}\label{eqn4}
L\lbrace\frac{d^\alpha f(t}{dt^\alpha}\rbrace=s^q L\lbrace f(t)\rbrace
\end{equation}
However, the transfer function F(s) for the fractional integral operator of order $\alpha$ can therefore be expressed as $F(s)=\frac{1}{s^q}$ in the frequency domain.With approximation approach, the errors of 2 dB have been deduce for fractional integral operator $\frac{1}{s^\alpha}$ \cite{Ahmad}.

\subsection{State dependent noise}\label{section:noise}
\begin{figure}
\centering\includegraphics[scale=0.55]{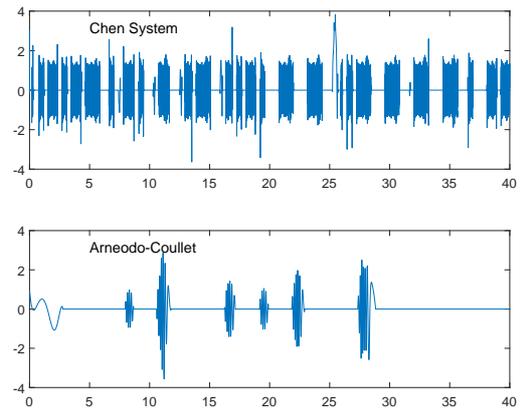}
\caption{State dependent noise for (top figure) the Chen system with system parameters $(a,b,c) = (40, 3, 28)$ using fractional order 0.9  (bottom figure) the Arneodo-Coullet system with parameters $(e,f)=(6 5.0)$ using fractional order 0.9}\label{fig0}.
\end{figure}

We define a state-dependent noise as feedback noise arising from one or more components of the system.   Consider a system defined as $f(x,y,z)$.  If a fraction of the output $y$ is made to interfere with the input $x$, a state-dependent noise has been introduced. This state-dependent noise can be a linear or nonlinear function of the state variables.  Box and Muller \cite{box1958gep} proposed that using Equation \ref{noise}, two random number streams ($U_1$ and $U_2$) can be used to generate standard random numbers.  In this study, we consider two systems $f(x,y,z)$ where the state $z$ is perturbed with feedback noise which is a nonlinear function of $x = U_1$ and $y = U_2$ states.
\begin{equation}\label{noise}
    \xi(U_1,U_2) = d\left(\sqrt{(-2\log U_1)}\cos2\pi U_2 \right)
\end{equation}
where $d$ is the noise strength. If the states of the system are independently chaotic, the noise $\xi(x,y)$ can be regarded as a chaotic noise. The statistical properties of $\xi(U_1,U_2)$ are shown in Table \ref{tab1} and the time series in Figure \ref{fig0}.

\begin{table*}[htb]
\centering
\caption{Statistical properties of state-dependent noise considered.}\label{tab1}
\begin{tabular}{lrrrrrr}
 &\multicolumn{3}{c}{Chen System}             &  \multicolumn{3}{c}{Arneodo-Coullet System }      \\
            & $U_1$           & $U_2$           & $\xi(U_1,U_2)$                     & $U_1$           & $U_2$           &$\xi(U_1,U_2)$          \\
Mean        & 0.8683  & 0.8808  & -0.0042             & 1.8461  & 0.0809  & 0.0051 \\
Median      & 1.6761  & 1.8132  & 0.3603             & 1.9403  & 0.3346  & 0.3774 \\
STD         & 7.6955  & 8.3352  & 1.6158            & 0.8544  & 1.7650   & 0.9030 \\
Skewness    & -0.1833 & -0.1776 & 0.0524            & -0.3723  & -0.4250 & 0.1719
\end{tabular}
\end{table*}

\subsection{Systems}
The Chen \cite{CHEN1999} and Arneodo-Coullet \cite{arneodo1981} chaotic systems were considered in this study.  The integer order Chen system is defined as
\begin{equation}\label{chenlee}
    \begin{split}
      \dot{x} &= a(y-x) \\
      \dot{y} &= (c - a)x - xz + cy\\
      \dot{z} &= (xy - bz)
    \end{split}
\end{equation}
This system has been shown to be chaotic when $a=35$, $b=3$, and $c=28$.  Chen system is related to but not topological equivalent to either the Lorenz  \cite{lorenz1963} or Rossler \cite {rossler1976equation} chaotic systems.

The Arneodo-Coullet chaotic system is defined as
\begin{equation}\label{arneodo}
    \begin{split}
      \dot{x} &= y \\
      \dot{y} &= z\\
      \dot{z} &= ex - fy - z - x^3
    \end{split}
\end{equation}
The fractional order form of the Chen and Arneodo-Coullet systems with continuous time, state-space noise (Section \ref{section:noise}) are given in Equations \ref{chenlee1} and \ref{arneodo1} respectively.
\begin{equation}\label{chenlee1}
    \begin{split}
      \frac{d^\alpha x}{dt^\alpha} &= a(y-x) \\
      \frac{d^\alpha y}{dt^\alpha}  &= (c - a)x - xz + cy\\
      \frac{d^\alpha y}{dt^\alpha}  &= (xy - bz)  + \xi(x,y)
    \end{split}
\end{equation}

\begin{equation}\label{arneodo1}
    \begin{split}
      \frac{d^\alpha x}{dt^\alpha}  &= y \\
      \frac{d^\alpha y}{dt^\alpha}  &= z\\
      \frac{d^\alpha z}{dt^\alpha} &= ex - fy - z - x^3 + \xi(x,y)
    \end{split}
\end{equation}
The constants are given as $a=35$, $b=3$, $c=28$, $e=5.5$ and $f=5.0$ for the two systems.  The circuit realization and phase space for the two fractional order systems are shown in \ref{appendix}

\section{Results and Discussion}
\begin{figure*}
\centering\includegraphics[scale=0.5]{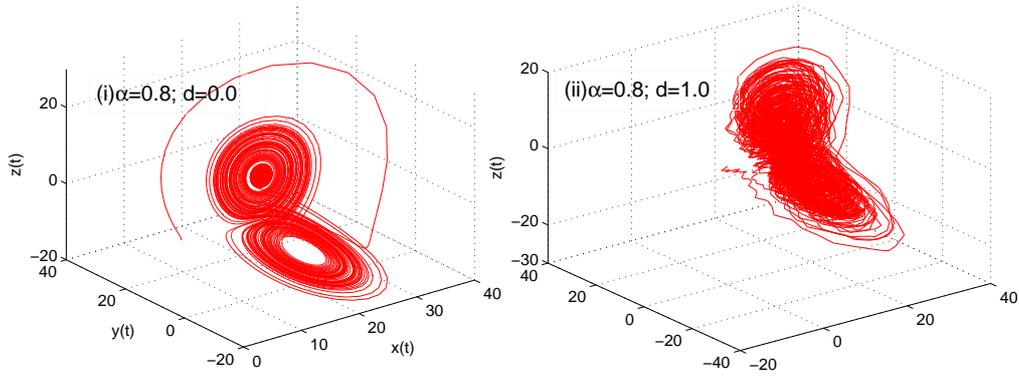}
\caption{3D Phase space representation of system \ref{chenlee1} with fractional order $\alpha = 0.8$ and noise level, $d$ (a) 0.0; and (b) 1.0}\label{fig2}.
\end{figure*}

\begin{figure*}
\centering\includegraphics[scale=0.5]{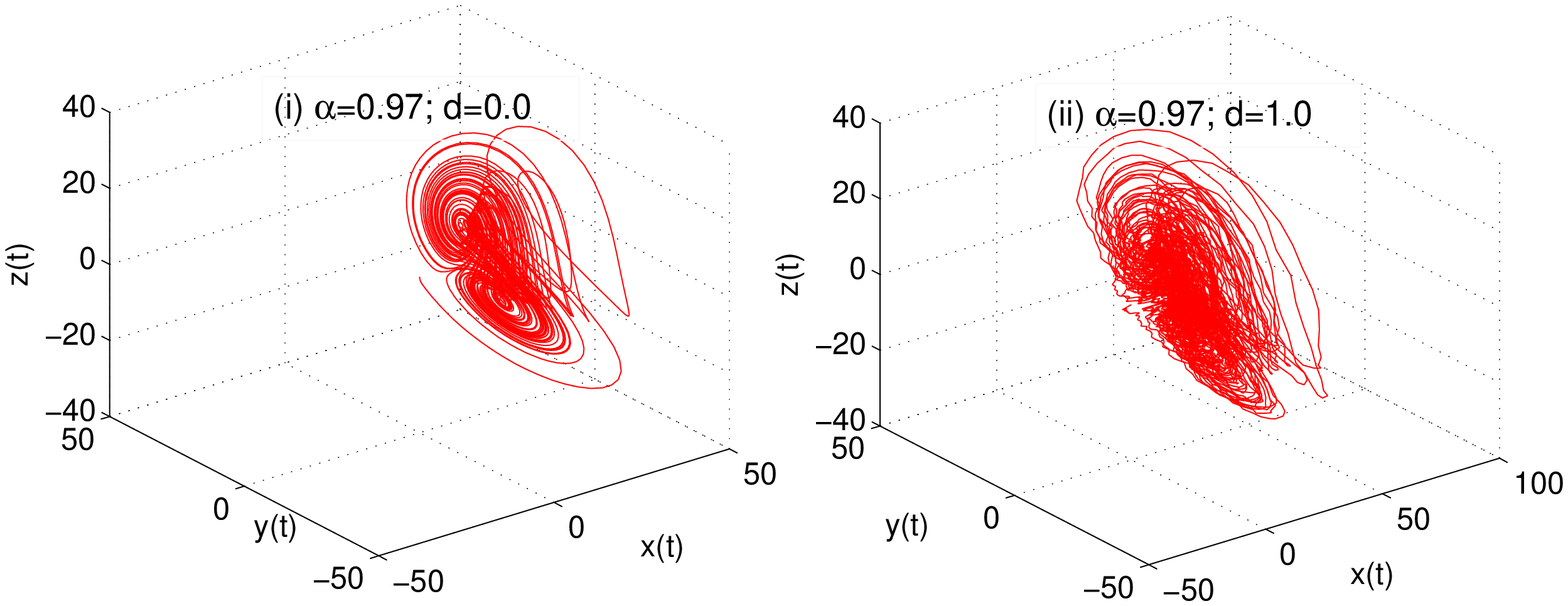}
\caption{3D Phase space representation of system \ref{arneodo1} with fractional order $\alpha = 0.97$ and noise level, $d$ (a) 0.0; and (b) 1.0}\label{fig3}.
\end{figure*}

Numerical simulations of systems (\ref{chenlee1}) and (\ref{arneodo1}) using different fractional order, $\alpha$ and noise level were implemented.  Using fractional orders of $0.8$ (Figure \ref{fig2}) and $0.97$ (Figure \ref{fig3}), the 3D phase space were found to experience distortion.  With a fractional order of $0.97$, the bifurcation diagram of Chen system was found to undergo significant changes.  Without the influence of state-dependent noise, the system is chaotic in the region $36<a<42.5$, however, the introduction of state-dependent noise with intensity $d=1.0$, this regime was found in the range $34<a<44$.  The periodic regime ($42.5<a<45$) and chaotic regime ($45<a<49$) were not observed when state dependent noise was introduced into the system.  Similar distortions were found in the phase space (Figure \ref{fig3}) and bifurcation (Figure \ref{fig5}) diagram of the Arneodo-Colluet systems.  In the bifurcation diagram of the Arneodo-Colluet system, the period doubling route to chaos was preserved but distorted.  The introduction of of state-dependent noise was found to suppress chaos in the two systems considered.  For increased values of $d$, it is expected that the chaotic state of the system will be destroyed.  The inhibition of the period doubling cascade by noise has also been reported in \cite{gao}.

\begin{figure}
\centering\includegraphics[width=0.45\textwidth]{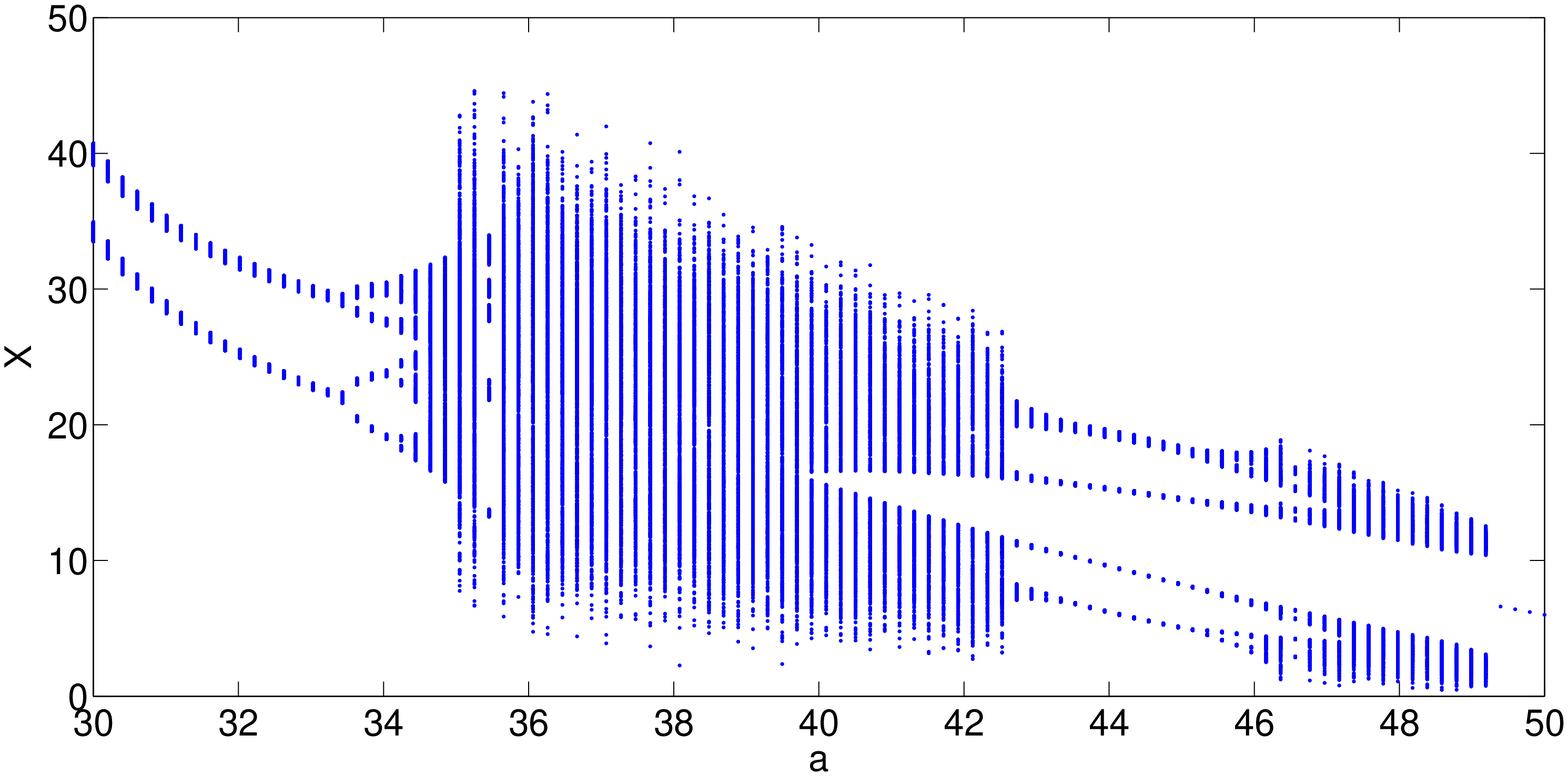}
\centering\includegraphics[width=0.45\textwidth]{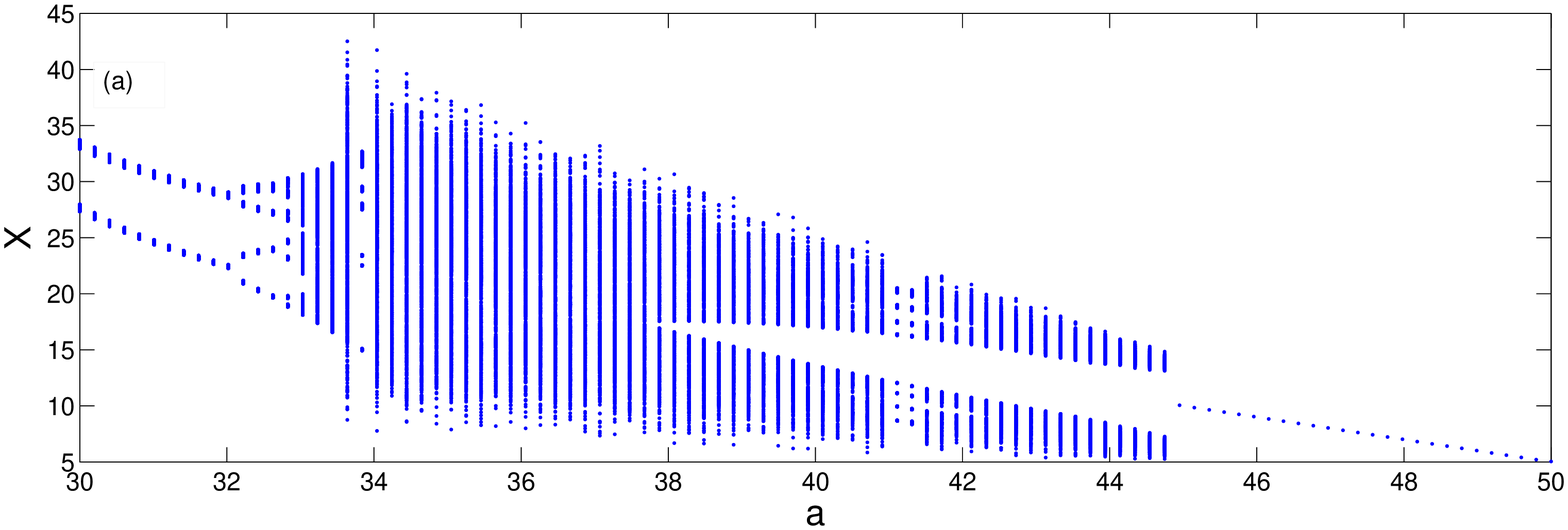}
\caption{Bifurcation diagram of the Chen system (Equation \ref{chenlee1}) with fractional order $\alpha = 0.97$ and noise level, $d$ (top) 0.0; and (bottom) 1.0}\label{fig4}.
\end{figure}

\begin{figure}
\centering\includegraphics[width=0.45\textwidth]{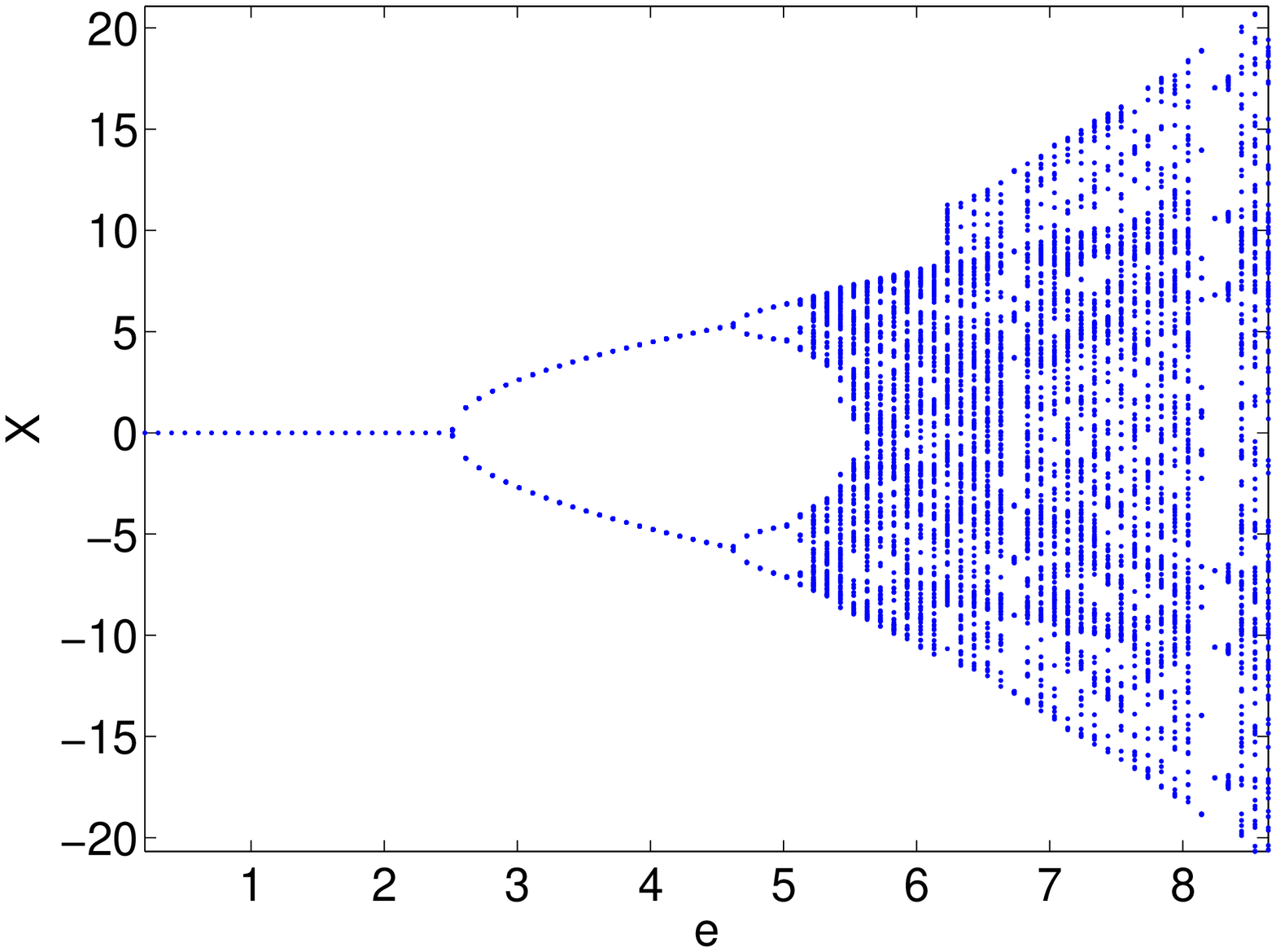}
\centering\includegraphics[width=0.45\textwidth]{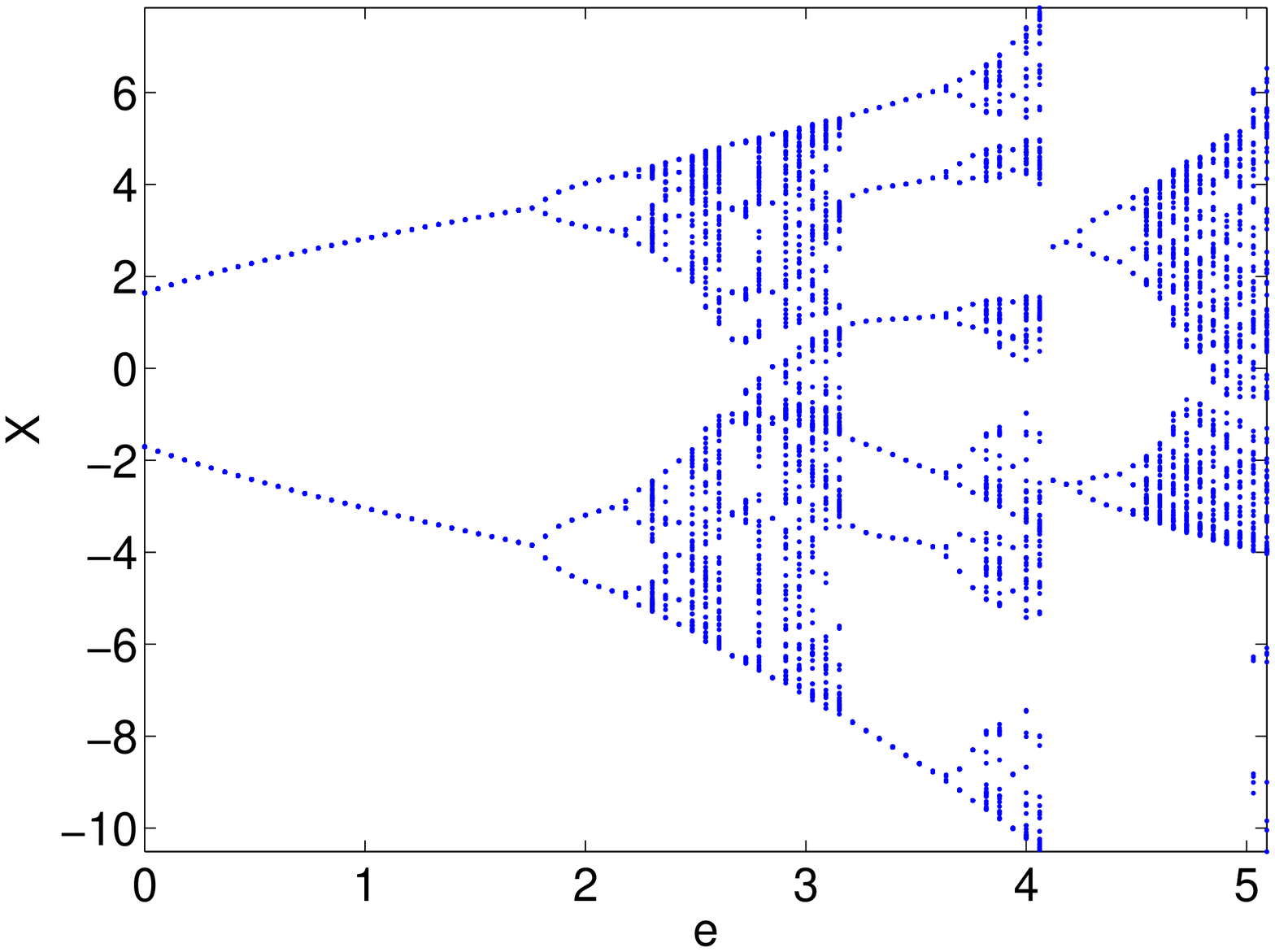}
\caption{Bifurcation diagram of the Arneodo-Colluet system (Equation \ref{arneodo1}) with fractional order $\alpha = 0.97$ and noise level, $d$ (top) 0.0; and (bottom) 0.5}\label{fig5}.
\end{figure}

\section{Conclusion}
In this work, we have considered the effect of state dependent noise on the Chen and Arneodo-Coullet chaotic systems.  The noise considered are dependent on feedback from other states of the system as against coloured noise.  Hence, they can be regarded as intrinsic noise.   The overall effect of state-dependent noise on the system studied is the suppression of chaos.  Thus, this approach will prove useful in the understanding of nonlinearity in systems susceptible to   interference from other states of the system.  The noise proposed in this study can be used as a convenient way to control chaos in a system using the system variables.


\appendix

\section{Circuit diagrams and simulation}\label{appendix}
\begin{figure*}
\centering\includegraphics[scale=0.2]{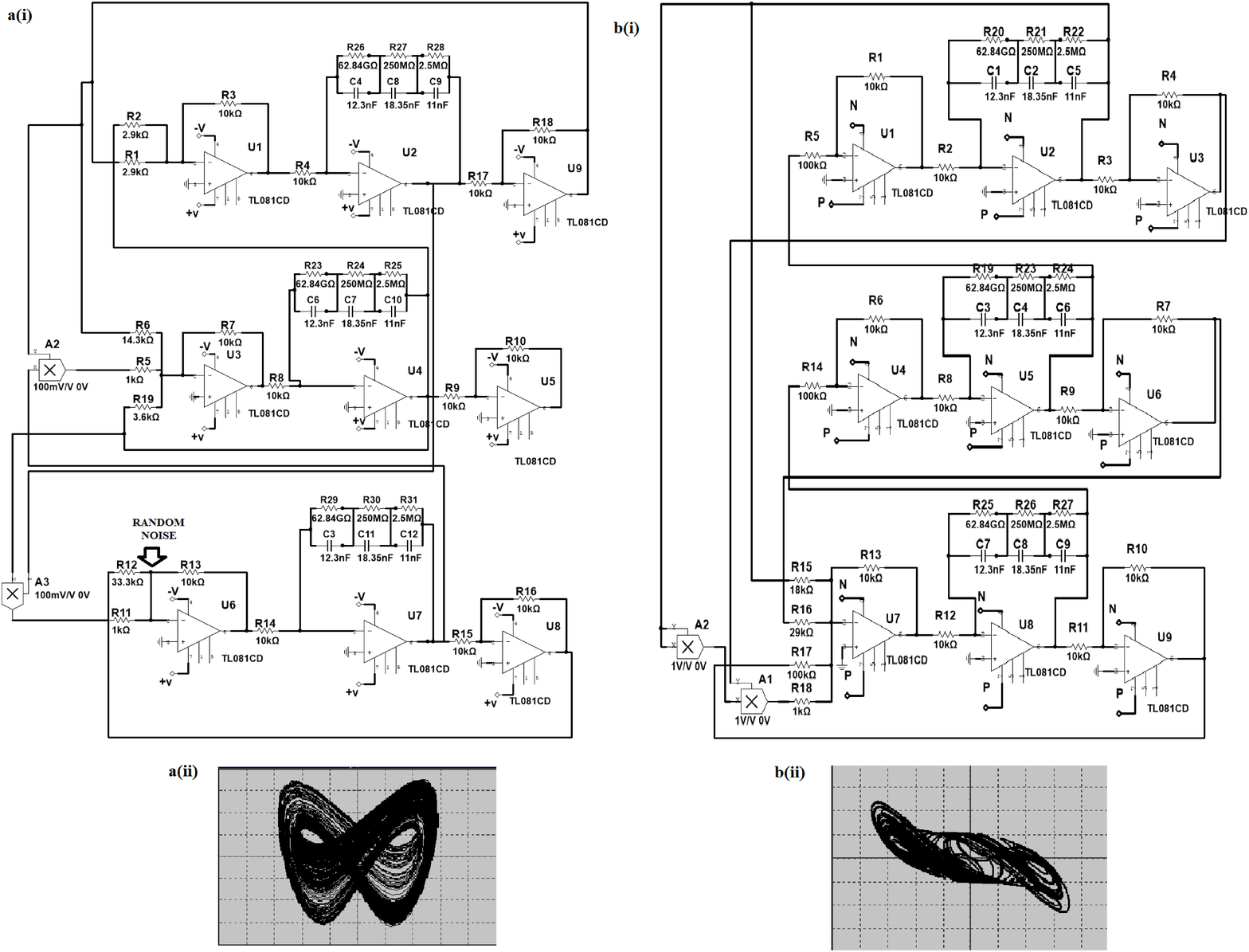}
\caption{Circuit diagram and phase space representation for the Chen (Equation \ref{chenlee1}) and Arneodo-Colluet (Equation \ref{arneodo1}) systems}\label{fig1}.
\end{figure*}

The circuit representation and resultant phase space are shown here.  The analog implementation of the two proposed systems can likewise  be generated with possibility of injecting Pseudo-random noise generator. The electronic circuits consists of chain fractance (combined capacitor and resistors), operational amplifiers (TL081CD), multipliers $(A_s)$, resistors and capacitors with power supply unit. Figure \ref{fig1} display the analog circuit realization of chaotic Chen and Arneodo-Coullet system  as well as the phase space representation.

\balance


\bibliographystyle{spphys}

\end{document}